\documentclass[preprint]{aastex}
\newcommand{\etal}{et al.}

\shorttitle{SDSS DR3}
\shortauthors{Abazajian \etal}
\begin{document}
\title{The Third Data Release of the Sloan Digital Sky Survey}
\author{
Kevork Abazajian\altaffilmark{\ref{LANLtheory}},
Jennifer K. Adelman-McCarthy\altaffilmark{\ref{Fermilab}},
Marcel A. Ag\"ueros\altaffilmark{\ref{Washington}},
Sahar S. Allam\altaffilmark{\ref{Fermilab}},
Kurt S. J. Anderson\altaffilmark{\ref{APO},\ref{NMSU}},
Scott F. Anderson\altaffilmark{\ref{Washington}},
James Annis\altaffilmark{\ref{Fermilab}},
Neta A. Bahcall\altaffilmark{\ref{Princeton}},
Ivan K. Baldry\altaffilmark{\ref{JHU}},
Steven Bastian\altaffilmark{\ref{Fermilab}},
Andreas Berlind\altaffilmark{\ref{Chicago},\ref{CfCP},\ref{NYU}},
Mariangela Bernardi\altaffilmark{\ref{Pitt},\ref{CMU}},
Michael R. Blanton\altaffilmark{\ref{NYU}},
John J. Bochanski Jr.\altaffilmark{\ref{Washington}},
William N. Boroski\altaffilmark{\ref{Fermilab}},
Howard J. Brewington\altaffilmark{\ref{APO}},
John W. Briggs\altaffilmark{\ref{NMIT}},
J. Brinkmann\altaffilmark{\ref{APO}},
Robert J. Brunner\altaffilmark{\ref{Illinois}},
Tam\'as Budav\'ari\altaffilmark{\ref{JHU}},
Larry N. Carey\altaffilmark{\ref{Washington}},
Francisco J. Castander\altaffilmark{\ref{Barcelona}},
A. J. Connolly\altaffilmark{\ref{Pitt}},
Kevin R. Covey\altaffilmark{\ref{Washington}},
Istv\'an Csabai\altaffilmark{\ref{Eotvos},\ref{JHU}},
Julianne J. Dalcanton\altaffilmark{\ref{Washington}},
Mamoru Doi\altaffilmark{\ref{IoAUT}},
Feng Dong\altaffilmark{\ref{Princeton}},
Daniel J. Eisenstein\altaffilmark{\ref{Arizona}},
Michael L. Evans\altaffilmark{\ref{Washington}},
Xiaohui Fan\altaffilmark{\ref{Arizona}},
Douglas P. Finkbeiner\altaffilmark{\ref{Princeton},\ref{Russell}},
Scott D. Friedman\altaffilmark{\ref{STScI}},
Joshua A. Frieman\altaffilmark{\ref{Fermilab},\ref{Chicago},\ref{CfCP}},
Masataka Fukugita\altaffilmark{\ref{ICRRUT}},
Bruce Gillespie\altaffilmark{\ref{APO}},
Karl Glazebrook\altaffilmark{\ref{JHU}},
Jim Gray\altaffilmark{\ref{Microsoft}},
Eva K. Grebel\altaffilmark{\ref{MPIA}},
James E. Gunn\altaffilmark{\ref{Princeton}},
Vijay K. Gurbani\altaffilmark{\ref{Fermilab},\ref{Lucent2}},
Patrick B. Hall\altaffilmark{\ref{Princeton},\ref{York}},
Masaru Hamabe\altaffilmark{\ref{JapanWomen}},
Daniel Harbeck\altaffilmark{\ref{Wisconsin}},
Frederick H. Harris\altaffilmark{\ref{NOFS}},
Hugh C. Harris\altaffilmark{\ref{NOFS}},
Michael Harvanek\altaffilmark{\ref{APO}},
Suzanne L. Hawley\altaffilmark{\ref{Washington}},
Jeffrey Hayes\altaffilmark{\ref{Catholic}},
Timothy M. Heckman\altaffilmark{\ref{JHU}},
John S. Hendry\altaffilmark{\ref{Fermilab}},
Gregory S. Hennessy\altaffilmark{\ref{USNO}},
Robert B. Hindsley\altaffilmark{\ref{NRL}},
Craig J. Hogan\altaffilmark{\ref{Washington}},
David W. Hogg\altaffilmark{\ref{NYU}},
Donald J. Holmgren\altaffilmark{\ref{Fermilab}},
Jon A. Holtzman\altaffilmark{\ref{NMSU}},
Shin-ichi Ichikawa\altaffilmark{\ref{NAOJ}},
Takashi Ichikawa\altaffilmark{\ref{TohokuU}},
\v{Z}eljko Ivezi\'{c}\altaffilmark{\ref{Princeton}},
Sebastian Jester\altaffilmark{\ref{Fermilab}},
David E. Johnston\altaffilmark{\ref{Princeton}},
Anders M. Jorgensen\altaffilmark{\ref{LANL}},
Mario Juri\'{c}\altaffilmark{\ref{Princeton}},
Stephen M. Kent\altaffilmark{\ref{Fermilab}},
S. J. Kleinman\altaffilmark{\ref{APO}},
G. R. Knapp\altaffilmark{\ref{Princeton}},
Alexei Yu. Kniazev\altaffilmark{\ref{MPIA}},
Richard G. Kron\altaffilmark{\ref{Chicago},\ref{Fermilab}},
Jurek Krzesinski\altaffilmark{\ref{APO},\ref{MSO}},
Donald Q. Lamb\altaffilmark{\ref{Chicago},\ref{EFI}},
Hubert Lampeitl\altaffilmark{\ref{Fermilab}},
Brian C. Lee\altaffilmark{\ref{LBL}},
Huan Lin\altaffilmark{\ref{Fermilab}},
Daniel C. Long\altaffilmark{\ref{APO}},
Jon Loveday\altaffilmark{\ref{Sussex}},
Robert H. Lupton\altaffilmark{\ref{Princeton}},
Ed Mannery\altaffilmark{\ref{Washington}},
Bruce Margon\altaffilmark{\ref{STScI}},
David Mart\'{\i}nez-Delgado\altaffilmark{\ref{MPIA}},
Takahiko Matsubara\altaffilmark{\ref{Nagoya}},
Peregrine M. McGehee\altaffilmark{\ref{NMSU},\ref{LANL2}},
Timothy A. McKay\altaffilmark{\ref{Michigan}},
Avery Meiksin\altaffilmark{\ref{Edinburgh}},
Brice M\'enard\altaffilmark{\ref{IAS}},
Jeffrey A. Munn\altaffilmark{\ref{NOFS}},
Thomas Nash\altaffilmark{\ref{Fermilab}},
Eric H. Neilsen, Jr.\altaffilmark{\ref{Fermilab}},
Heidi Jo Newberg\altaffilmark{\ref{RPI}},
Peter R. Newman\altaffilmark{\ref{APO}},
Robert C. Nichol\altaffilmark{\ref{Portsmouth}},
Tom Nicinski\altaffilmark{\ref{Fermilab},\ref{CMCElectronics}},
Maria Nieto-Santisteban\altaffilmark{\ref{STScI}},
Atsuko Nitta\altaffilmark{\ref{APO}},
Sadanori Okamura\altaffilmark{\ref{DoAUT}},
William O'Mullane\altaffilmark{\ref{JHU}},
Russell Owen\altaffilmark{\ref{Washington}},
Nikhil Padmanabhan\altaffilmark{\ref{Princetonphys}},
George Pauls\altaffilmark{\ref{Princeton}},
John Peoples\altaffilmark{\ref{Fermilab}},
Jeffrey R. Pier\altaffilmark{\ref{NOFS}},
Adrian C. Pope\altaffilmark{\ref{JHU}},
Dimitri Pourbaix\altaffilmark{\ref{Princeton},\ref{Bruxelles}},
Thomas R. Quinn\altaffilmark{\ref{Washington}},
M. Jordan Raddick\altaffilmark{\ref{JHU}},
Gordon T. Richards\altaffilmark{\ref{Princeton}},
Michael W. Richmond\altaffilmark{\ref{RIT}},
Hans-Walter Rix\altaffilmark{\ref{MPIA}},
Constance M. Rockosi\altaffilmark{\ref{Lick}},
David J. Schlegel\altaffilmark{\ref{Princeton}},
Donald P. Schneider\altaffilmark{\ref{PSU}},
Joshua Schroeder\altaffilmark{\ref{Princeton},\ref{Colorado}},
Ryan Scranton\altaffilmark{\ref{Pitt}},
Maki Sekiguchi\altaffilmark{\ref{JPG}},
Erin Sheldon\altaffilmark{\ref{Chicago},\ref{CfCP}},
Kazu Shimasaku\altaffilmark{\ref{DoAUT}},
Nicole M. Silvestri\altaffilmark{\ref{Washington}},
J. Allyn Smith\altaffilmark{\ref{Wyoming},\ref{LANL}},
Vernesa Smol\v{c}i\'{c}\altaffilmark{\ref{Zagreb}},
Stephanie A. Snedden\altaffilmark{\ref{APO}},
Albert Stebbins\altaffilmark{\ref{Fermilab}},
Chris Stoughton\altaffilmark{\ref{Fermilab}},
Michael A. Strauss\altaffilmark{\ref{Princeton}},
Mark SubbaRao\altaffilmark{\ref{Chicago},\ref{Adler}},
Alexander S. Szalay\altaffilmark{\ref{JHU}},
Istv\'an Szapudi\altaffilmark{\ref{Hawaii}},
Paula Szkody\altaffilmark{\ref{Washington}},
Gyula P. Szokoly\altaffilmark{\ref{MPIEP}},
Max Tegmark\altaffilmark{\ref{Penn}},
Luis Teodoro\altaffilmark{\ref{LANLtheory}},
Aniruddha R. Thakar\altaffilmark{\ref{JHU}},
Christy Tremonti\altaffilmark{\ref{Arizona}},
Douglas L. Tucker\altaffilmark{\ref{Fermilab}},
Alan Uomoto\altaffilmark{\ref{JHU},\ref{CarnegieObs}},
Daniel E. Vanden Berk\altaffilmark{\ref{PSU}},
Jan Vandenberg\altaffilmark{\ref{JHU}},
Michael S. Vogeley\altaffilmark{\ref{Drexel}},
Wolfgang Voges\altaffilmark{\ref{MPIEP}},
Nicole P. Vogt\altaffilmark{\ref{NMSU}},
Lucianne M. Walkowicz\altaffilmark{\ref{Washington}},
Shu-i Wang\altaffilmark{\ref{CMCElectronics}},
David H. Weinberg\altaffilmark{\ref{OSU}},
Andrew A. West\altaffilmark{\ref{Washington}},
Simon D.M. White\altaffilmark{\ref{MPA}},
Brian C. Wilhite\altaffilmark{\ref{Chicago}},
Yongzhong Xu\altaffilmark{\ref{LANLtheory}},
Brian Yanny\altaffilmark{\ref{Fermilab}},
Naoki Yasuda\altaffilmark{\ref{ICRRUT}},
Ching-Wa Yip\altaffilmark{\ref{Pitt}},
D. R. Yocum\altaffilmark{\ref{Fermilab}},
Donald G. York\altaffilmark{\ref{Chicago},\ref{EFI}},
Idit Zehavi\altaffilmark{\ref{Arizona}},
Stefano Zibetti\altaffilmark{\ref{MPA}},
Daniel B. Zucker\altaffilmark{\ref{MPIA}}
}

\altaffiltext{1}{
Theoretical Division, MS B285, Los Alamos National Laboratory, Los Alamos, NM 87545.
\label{LANLtheory}}

\altaffiltext{2}{
Fermi National Accelerator Laboratory, P.O. Box 500, Batavia, IL 60510.
\label{Fermilab}}

\altaffiltext{3}{
Department of Astronomy, University of Washington, Box 351580, Seattle, WA
98195.
\label{Washington}}

\altaffiltext{4}{
Apache Point Observatory, P.O. Box 59, Sunspot, NM 88349.
\label{APO}}

\altaffiltext{5}{
Department of Astronomy, MSC 4500, New Mexico State University,
P.O. Box 30001, Las Cruces, NM 88003.
\label{NMSU}}

\altaffiltext{6}{
Department of Astrophysical Sciences, Princeton University, Princeton, NJ
08544.
\label{Princeton}}

\altaffiltext{7}{
Center for Astrophysical Sciences, Department of Physics and Astronomy, Johns
Hopkins University, 3400 North Charles Street, Baltimore, MD 21218. 
\label{JHU}}

\altaffiltext{8}{
Department of Astronomy and Astrophysics, University of Chicago, 5640 South
Ellis Avenue, Chicago, IL 60637.
\label{Chicago}}

\altaffiltext{9}{
Center for Cosmological Physics, The University of Chicago,
5640 South Ellis Avenue, Chicago, IL 60637.
\label{CfCP}}

\altaffiltext{10}{
Center for Cosmology and Particle Physics,
Department of Physics,
New York University,
4 Washington Place,
New York, NY 10003.
\label{NYU}}

\altaffiltext{11}{
Department of Physics and Astronomy, University of Pittsburgh, 3941 O'Hara
Street, Pittsburgh, PA 15260.
\label{Pitt}}

\altaffiltext{12}{
Department of Physics, Carnegie Mellon University, 5000 Forbes Avenue,
Pittsburgh, PA 15213. 
\label{CMU}}

\altaffiltext{13}{
Madgalena Ridge Observatory
New Mexico Institute of Technology
801 Leroy Place
Socorro, NM  87801.
\label{NMIT}}

\altaffiltext{14}{
Department of Astronomy
University of Illinois
1002 West Green Street, Urbana, IL 61801.
\label{Illinois}}

\altaffiltext{15}{Institut d'Estudis Espacials de Catalunya/CSIC, Gran Capit\'a 2-4,
E-08034 Barcelona, Spain.
\label{Barcelona}}

\altaffiltext{16}{
Department of Physics of Complex Systems, E\"{o}tv\"{o}s Lor\'and University, Pf.\ 32,
H-1518 Budapest, Hungary.
\label{Eotvos}}

\altaffiltext{17}{Institute of Astronomy and Research Center for the
  Early Universe, School
of Science, University of Tokyo,
 2-21-1 Osawa, Mitaka, Tokyo 181-0015, Japan.
\label{IoAUT}}

\altaffiltext{18}{
Steward Observatory, 933 North Cherry Avenue, Tucson, AZ 85721.
\label{Arizona}}

\altaffiltext{19}{
Russell-Cotsen Fellow
\label{Russell}}

\altaffiltext{20}{
Space Telescope Science Institute, 3700 San Martin Drive, Baltimore, MD
21218.
\label{STScI}}

\altaffiltext{21}{Institute for Cosmic Ray Research, University of Tokyo, 5-1-5 Kashiwa,
 Kashiwa City, Chiba 277-8582, Japan.
\label{ICRRUT}}

\altaffiltext{22}{
Microsoft Research, 455 Market Street, Suite 1690, San Francisco, CA 94105.
\label{Microsoft}}

\altaffiltext{23}{
Max-Planck-Institut f\"ur Astronomie, K\"onigstuhl 17, D-69117 Heidelberg,
Germany.
\label{MPIA}}

\altaffiltext{24}{
Lucent Technologies, 2000 Lucent Lane, Naperville, IL 60566.
\label{Lucent2}}

\altaffiltext{25}{
Dept. of Physics \& Astronomy,
York University,
4700 Keele St.,
Toronto, ON, M3J 1P3,
Canada
\label{York}}

\altaffiltext{26}{
Department of Mathematical and Physical Sciences,
Japan Women's University
2-8-1 Mejirodai, Bunkyo, Tokyo 112-8681, Japan.
\label{JapanWomen}}

\altaffiltext{27}{
Department of Astronomy, University of Wisconsin, 475 North Charter
Street, Madison, WI 53706.
\label{Wisconsin}}

\altaffiltext{28}{
US Naval Observatory, Flagstaff Station, P.O. Box 1149, Flagstaff, AZ
86002-1149.
\label{NOFS}}

\altaffiltext{29}{
Institute for Astronomy and Computational Sciences
     Physics Department
     Catholic University of America
     Washington DC 20064
\label{Catholic}}

\altaffiltext{30}{
US Naval Observatory, 3540 Massachusetts Avenue NW, Washington, DC 20392.
\label{USNO}}

\altaffiltext{31}{
Code 7215, Remote Sensing Division
Naval Research Laboratory
4555 Overlook Avenue SW
Washington, DC 20392
\label{NRL}}

\altaffiltext{32}{National Astronomical Observatory, 2-21-1 Osawa, Mitaka, Tokyo 181-8588,
Japan.
\label{NAOJ}}

\altaffiltext{33}{Astronomical Institute, Tohoku University, Aramaki, Aoba, Sendai 980-8578,
Japan.
\label{TohokuU}}

\altaffiltext{34}{
ISR-4, MS D448, Los Alamos National Laboratory, P.O.Box 1663, Los Alamos, NM 87545.
\label{LANL}}

\altaffiltext{35}{
Obserwatorium Astronomiczne na Suhorze, Akademia Pedogogiczna w
Krakowie, ulica Podchor\c{a}\.{z}ych 2,
PL-30-084 Krac\'ow, Poland.
\label{MSO}}

\altaffiltext{36}{
Enrico Fermi Institute, University of Chicago, 5640 South Ellis Avenue,
Chicago, IL 60637.
\label{EFI}}

\altaffiltext{37}{
Lawrence Berkeley National Laboratory, One Cyclotron Road,
Berkeley CA 94720-8160.
\label{LBL}}

\altaffiltext{38}{
Astronomy Centre, University of Sussex, Falmer, Brighton BN1 9QJ, UK. 
\label{Sussex}}

\altaffiltext{39}{
Department of Physics and Astrophysics,
 Nagoya University,
 Chikusa, Nagoya 464-8602,
 Japan.
\label{Nagoya}}

\altaffiltext{40}{
SNS-4, MS H820, Los Alamos National Laboratory, P.O.Box 1663, Los Alamos, NM 87545.
\label{LANL2}}

\altaffiltext{41}{
Department of Physics, University of Michigan, 500 East University Avenue, Ann
Arbor, MI 48109.
\label{Michigan}}

\altaffiltext{42}{
Institute for Astronomy,
Royal Observatory,
University of Edinburgh,
Blackford Hill,
Edinburgh EH9 3HJ,
UK.
\label{Edinburgh}}

\altaffiltext{43}{
Institute for Advanced Study, Olden Lane, Princeton, NJ 08540
\label{IAS}}

\altaffiltext{44}{
Department of Physics, Applied Physics, and Astronomy, Rensselaer
Polytechnic Institute, 110 Eighth Street, Troy, NY 12180. 
\label{RPI}}

\altaffiltext{45}{
Institute of Cosmology and Gravitation (ICG),
Mercantile House, Hampshire Terrace,
Univ. of Portsmouth, Portsmouth, PO1 2EG, UK
\label{Portsmouth}}

\altaffiltext{46}{
    CMC Electronics Aurora,
 84 N. Dugan Rd.
    Sugar Grove, IL 60554.
\label{CMCElectronics}}

\altaffiltext{47}{Department of Astronomy and Research Center for the Early Universe, 
University of Tokyo,
 7-3-1 Hongo, Bunkyo, Tokyo 113-0033, Japan.
\label{DoAUT}}

\altaffiltext{48}{
Joseph Henry Laboratories, Princeton University, Princeton, NJ
08544.
\label{Princetonphys}}

\altaffiltext{49}{
FNRS
Institut  d'Astronomie et d'Astrophysique,
 Universit\'e Libre de Bruxelles, CP. 226, Boulevard du Triomphe, B-1050
 Bruxelles, Belgium
\label{Bruxelles}}

\altaffiltext{50}{
Department of Physics, Rochester Institute of Technology, 84 Lomb Memorial
Drive, Rochester, NY 14623-5603.
\label{RIT}}

\altaffiltext{51}{
UCO/Lick Observatory, University of California, Santa Cruz, CA 95064
\label{Lick}}

\altaffiltext{52}{
Department of Astronomy and Astrophysics, 525 Davey Laboratory, Pennsylvania State
University, University Park, PA 16802.
\label{PSU}}

\altaffiltext{53}{
Center for Astrophysics and Space Astronomy, University of Colorado,
Boulder, CO 80309.
\label{Colorado}}

\altaffiltext{54}{
Japan Participation Group, 
c/o Institute for Cosmic Ray Research, University of Tokyo, 5-1-5
Kashiwa, Kashiwa City, Chiba 277-8582, Japan.
\label{JPG}}

\altaffiltext{55}{
Department of Physics and Astronomy, University of Wyoming, Laramie, WY 82071.
\label{Wyoming}}

\altaffiltext{56}{University of Zagreb, 
Department of Physics, Bijeni\v{c}ka cesta 32, 
10000 Zagreb, Croatia.
\label{Zagreb}}

\altaffiltext{57}{
Adler Planetarium and Astronomy Museum,
1300 Lake Shore Drive,
Chicago, IL 60605.
\label{Adler}}

\altaffiltext{58}{
Institute for Astronomy, 2680 Woodlawn Road, Honolulu, HI 96822.
\label{Hawaii}}

\altaffiltext{59}{
Max-Planck-Institut f\"ur extraterrestrische Physik, 
Giessenbachstrasse 1, D-85741 Garching, Germany.
\label{MPIEP}}

\altaffiltext{60}{
Department of Physics, University of Pennsylvania, Philadelphia, PA 19104.
\label{Penn}}

\altaffiltext{61}{
Observatories of the Carnegie Institution of Washington, 
813 Santa Barbara Street, 
Pasadena, CA  91101.
\label{CarnegieObs}}

\altaffiltext{62}{
Department of Physics, Drexel University, 3141 Chestnut Street, Philadelphia, PA 19104.
\label{Drexel}}

\altaffiltext{63}{
Department of Astronomy, Ohio State University, 140 West 8th Avenue, Columbus, OH 43210.
\label{OSU}}

\altaffiltext{64}{
Max Planck Institut f\"ur Astrophysik, Postfach 1, 
D-85748 Garching, Germany.
\label{MPA}}

\begin{abstract}
This paper describes the Third Data Release of the Sloan Digital Sky
Survey (SDSS).  This release, containing data taken up through June
2003, includes imaging data in five bands over 5282 deg$^2$,
photometric and astrometric catalogs of the 141 million objects
detected in these imaging data, and spectra of 528,640 objects
selected over 4188 deg$^2$.  The pipelines analyzing both images and
spectroscopy are unchanged from those used in our Second
Data Release.
 \end{abstract}

\keywords{Atlases---Catalogs---Surveys}

\section{Introduction}

The Sloan Digital Sky Survey (SDSS; York \etal\ 2000) is carrying out
an imaging and spectroscopic CCD survey of the sky at high Galactic
latitudes, using a dedicated wide-field 2.5m telescope at Apache Point
Observatory in South-East New Mexico.  The telescope saw first light in
May 1998, and following an extensive period of commissioning, formal
survey operations began in April 2000.  The resulting data have been
distributed to the public via web interfaces accessible from the SDSS
public web site\footnote{\tt http://www.sdss.org}, and have been
described in a series of papers:
\begin{itemize} 
  \item The Early Data Release, including data taken during
    commissioning, Stoughton \etal\ (2002; hereafter EDR paper),
    consisting of 462 deg$^2$ of imaging data and spectra of 54,000
    objects.  
   \item The First Data Release, Abazajian \etal\ (2003; hereafter DR1
     paper), consisting of imaging data over 2099 square degrees and
     spectra of 186,240 objects. 
   \item The Second Data Release, Abazajian \etal\ (2004; hereafter
     DR2 paper), consisting of imaging data over 3324 square degrees
     and spectra of 367,360 objects. 
\end{itemize}

These data allow investigations in all areas of optical astronomy,
from asteroids to quasars.  Among the papers which have appeared
since the Second Data Release are studies of the stellar masses of
galaxies (Brinchmann \etal\ 2004; Kauffmann \etal\ 2004), measurements
of the dark matter power spectrum from the Ly $\alpha$ forest
(McDonald \etal\ 2004) and corresponding constraints on cosmological
parameters (Seljak \etal\ 2004); studies of quasars (Lacy \etal\ 2004)
and galaxies (Hogg \etal\ 2004) by cross-correlating SDSS data with
the Spitzer First-Look Survey; the discovery of new structures in the
halo of M31 (Zucker \etal\ 2004ab); measurements of the detailed shape
of the galaxy correlation function and relation to halo occupation
models (Zehavi \etal\ 2004); and studies of populations of
stars at the bottom of the HR diagram (West \etal\ 2004; Knapp \etal\
2004). 

The SDSS imaging data are taken on photometric moonless nights (with
photometricity determined by an auxiliary telescope; Hogg \etal\ 2001)
of good
seeing with a wide-field imaging camera operating in drift-scan mode
(Gunn \etal\ 1998).  Six parallel {\em scanlines} on the sky, each
$13^\prime$ wide, are observed by each of the columns of CCDs.  Each
of the five rows of the CCDs in the camera is fronted by a different
filter, thus each scanline is observed in five filters, denoted (in
order of observation) $r\,i\,u\,z\,g$ (Fukugita \etal\ 1996; Gunn
\etal\ 1998; EDR paper).  The imaging data are processed by automated
software pipelines that measure the properties of all detected objects
(Lupton \etal\ 2001), and perform astrometric (Pier \etal\ 2003) and
photometric calibration, the latter to a set of standard stars
observed with the Photometric Telescope (Smith \etal\ 2003).  The
resulting object catalogs are used to 
select targets for spectroscopy, including the ``Main'' sample of
galaxies, magnitude-limited to $r_{Petrosian} = 17.77$ (Strauss \etal\
2002), a sample of luminous red ellipticals to $r_{Petrosian} = 19.5$
(Eisenstein \etal\ 2001; DR2 paper), quasar candidates selected by
their colors to $i = 19.1$ (for $z < 3$ candidates) and $i=20.2$ (for
higher-redshift candidates) (Richards \etal\ 2002), as well as a host
of additional targets, including optical counterparts to ROSAT X-ray
sources (Anderson \etal\ 2003), unusual stars, and calibration
observations (EDR paper).  All magnitude limits here are corrected for
Galactic extinction following Schlegel, Finkbeiner, \& Davis (1998).
The list of spectroscopic targets is 
distributed among a series of spectroscopic {\it tiles} of $3^\circ$
diameter (Blanton \etal\ 2003) to maximize observing
efficiency.  Each tile then forms the design for a spectroscopic
plate: holes are drilled in aluminum plates corresponding to the
position of each object for which spectra will be measured.  At the
telescope, optical fibers feeding a pair of double spectrographs are
plugged into each plate; the spectroscopic observations, carried out
under conditions not pristine enough for imaging, are typically 45
minutes per plate.  The spectra are wavelength- and flux-calibrated,
and run through an automatic pipeline to classify them and determine
redshifts (EDR paper).  

  The previous data release papers describe the quality of the data;
  the basic attributes of the data are given in
  Table~\ref{tab:characteristics}.  

\section{The Third Data Release}

The SDSS Third Data Release (DR3) consists of all survey-quality data
taken through June 2003 as part of the main SDSS survey.  The
footprints of the imaging and spectroscopic data are shown in
Figure~\ref{fig:skydist}.  The spectroscopic footprint is smaller:
because spectroscopic targets are chosen from the imaging data, the
spectroscopy always lags the imaging. As with previous data releases, DR3 does
{\em not} include repeat imaging scans (mostly on the Celestial
Equator in the Southern Galactic Cap; see Figure~\ref{fig:skydist}),
repeat observations of spectroscopic plates that have been 
observed more than once, or imaging or spectroscopic data taken
significantly outside the ellipse of the main survey footprint (as
described in York \etal\ 
2000).  However, six spectroscopic plates, and thirty-four square
degrees of imaging data in the Northern Galactic Cap in DR3 lie just
outside this ellipse. As in previous data releases, DR3
does include imaging data that overlap between adjacent runs. 

The SDSS has taken a
considerable amount of imaging data at low Galactic latitudes outside
the survey footprint; the subset of these data taken before Summer
2003 have been made publically available in a release separate from
DR3, and is described by Finkbeiner \etal\ (2004).

  As the survey has progressed, we have steadily improved the software
  used to process the imaging and spectroscopic data.  These changes
  are described in the DR1 and DR2 papers, and at the public SDSS web
  site.  Each subsequent release has incorporated all the data
  included in the previous release, necessitating a reprocessing of
  those data.  For DR3, however, we have made {\em no} changes in the
  processing software, and therefore the DR2 subset of the DR3 data
  are {\em identical} to the data already made public in 2004 March.
   There is one exception to this statement, namely that we have
   updated some incorrect spectral classifications in DR3.  We have
   also added several new auxiliary tables to the SDSS Catalog  
Archive Server (CAS) which will be useful for those using SDSS data.  
The Archive Introduction page in the Help menu on the CAS website\footnote{http://skyserver.sdss.org} describes the CAS data model.  It has a 
new link at the top for release-specific notes.

\subsection{Updated redshifts}
We have visually inspected the spectra of all objects which
  remained unclassified (spectral class UNKNOWN), and have updated the
  classification and redshift where appropriate.  Many of these
  objects are of low signal-to-noise ratio, or are unusual objects of
  various types, especially unusual quasars (see Hall \etal\ 2002).
  There are 477 objects whose classifications were updated in this
  effort, including 377 objects included in DR2. 

\subsection{New auxiliary tables}
\begin{figure}[t]\centering\includegraphics[width=12cm]{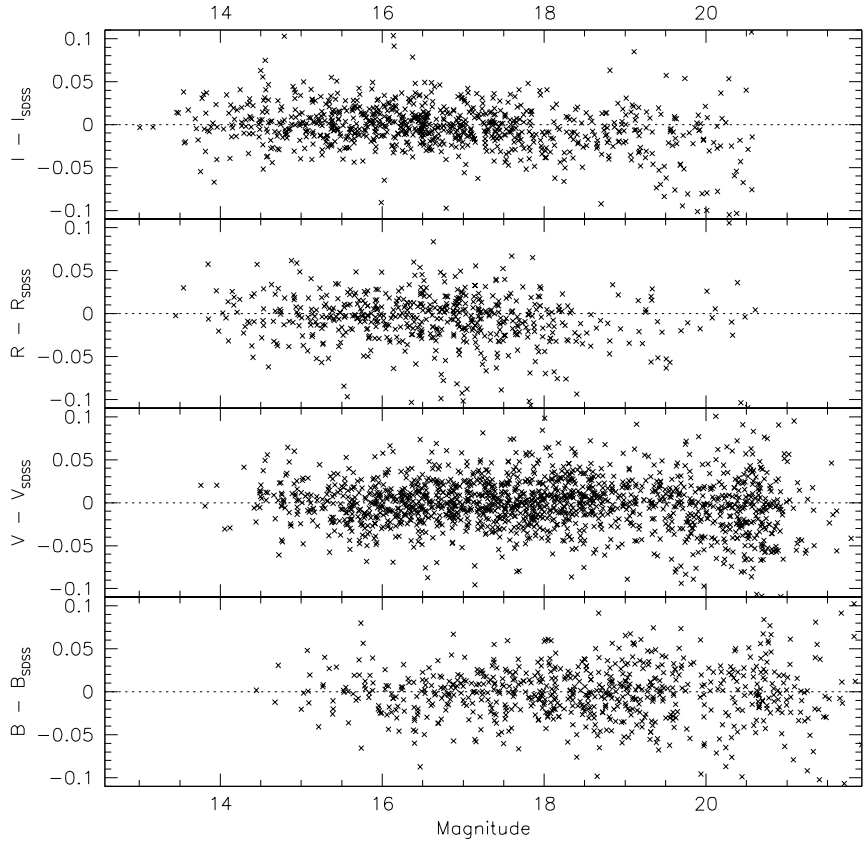}
\caption{A comparison of SDSS and Stetson (2000) photometry of stars.
  After finding the best-fit conversion from SDSS $ugriz$ photometry
  to the Kron-Cousins system used by Stetson, the residuals are found
  to be independent of magnitude (perhaps indicating that they are
  dominated by photometric calibration, and/or the uncertainties in
  the transformation between the two photometric systems), and have an
  rms of $\sim 2.5\%$.   
\label{fig:stetson}}\end{figure}

We have added a separate database table that includes stellar
  B, V, R, and I Kron-Cousins photometry with accuracy of 0.02 mag or
  better from Stetson (2000), as downloaded from the Canadian Astronomy Data Centre
  photometric standards website\footnote{\tt
    http://cadcwww.dao.nrc.ca/standards/}.     These data, when
  cross-matched with SDSS, allow SDSS photometry of stars to be
  compared with an external catalog.  After fitting for the conversion
  between the $ugriz$ SDSS photometric system to the Kron-Cousins
  photometry used by Stetson, we find the residuals shown in
  Figure~\ref{fig:stetson}; the rms scatter is of order 2.5\% in each
  band.  There is some evidence that this scatter is dominated by
  errors in the photometric calibration of the SDSS.  Further details
  will be given in 
  Holtzman \etal\ (in preparation).  We
  have also included data from the Third Reference Catalog of Galaxies
  (de Vaucouleurs \etal\ 1991), again to allow cross-reference to SDSS
  data. Please see the CAS Archive Introduction page on the SkyServer
  website's Help page for more information about the Stetson and RC3
  databases.

\subsection{Imaging quality measures on a field-by-field basis}

As part of quality assurance of the SDSS data, each of the $\sim
  200,000$ $10^\prime \times 13^\prime$ fields within DR3 is assigned a
  quality flag, {\tt FieldQAll}, which is now made available in a
  separate table,   entitled {\em RunQA} in both the Catalog Archive
  Server and Data Archive Server.  This flag is based on five attributes:  
\begin{itemize} 
\item The seeing in the $r$ band. 
\item The mean offset between the $7^{\prime\prime}$ aperture magnitude,
and the Point Spread Function (PSF) magnitude for bright stars on the
frame.  The accuracy of the
  results of the photometric pipeline are critically dependent on a
  correct model for the PSF, and the aperture minus PSF magnitude is
  an excellent diagnostic for problems in the PSF determination.  This
  magnitude difference typically is large under conditions in which
  the seeing is varying rapidly (see the discussion in Ivezi\'c \etal\ 2004).
\item Systematic offsets of the stellar locus in color space, and/or
  increased width of the stellar locus. 
As described in Fan (1999), Finlator \etal\ (2000),
  Helmi 
  \etal\ (2003), and many other papers, stars in the SDSS photometric
  system follow a narrow locus in color-color diagrams.  One can
  define a series of four {\em principal colors} from fits to the
  stellar locus, which are linear transformations of the SDSS
  magnitudes which empirically are essentially constant over the sky
  (after correcting for foreground reddening following Schlegel \etal\
  1998).  That is, systematic deviations of these principal colors as
  small as 1\% could indicate problems in the data.  These principal
  colors are defined as: 
\begin{itemize} 
\item The $s$ color, $s=-0.249u+0.794g-0.555r+0.234$, which is
  perpendicular to the stellar locus in the $u-g, g-r$ diagram. 
\item The $w$ color, $w=-0.227g+0.792r-0.567i+0.050$, which is
  perpendicular to the blue branch of the stellar locus in the $r-i,
  g-r$ diagram.  
\item The $x$ color, $x=0.707g-0.707r-0.983$, which is perpendicular
  to the red branch of the stellar locus in the $r-i, g-r$ diagram. 
\item The $y$ color, $y=-0.270r+0.800i-0.534z+0.059$, which is
  perpendicular to the stellar locus in the $r-i, i-z$ diagram. 
\end{itemize}

These principal colors are measured directly from the stars
  in running bins four fields wide.  Deviations from the global mean
  principal colors are indications of photometric errors, especially
  those due to photometric calibration.  Scatter around the principal
  colors within a single bin (i.e., a broad stellar locus) is another
  indication of poor data.  For the present, we use the so-called $s$
  color in our overall quality assessment of each field (although
  statistics on all   four colors are available in the {\em RunQA}
  table). In future work, we plan to incorporate information about all
  four colors in the field quality. 
 \item Finally, the processing of the data itself can indicate
  problems.  As described in \S 4.6 of the EDR paper (\S 4.6), a field can be
  given a so-called operational database quality of {\tt BAD}, {\tt
    MISSING}, or {\tt HOLE}, often because the 
  data corresponding to the field is particularly poor, or the
  photometric pipeline timed out on processing it (e.g., because of
  the presence of a naked-eye star in the field).  
\end{itemize}

Based on these quantities, we assign {\tt FieldQAll} for each field,
as follows:
\begin{itemize}
  \item By default, the field in question is assigned {\tt FieldQAll=
    ACCEPTABLE} (listed in the {\it RunQA} table numerically as {\bf 1}).  
  \item If the absolute value of the median PSF-aperture difference is
    greater than 0.05 magnitudes in any of the five bands, or if the
    absolute value of the $s$ color median is larger than 0.05
    magnitudes, or if the $s$ color distribution width is 2.5 times
    larger than its median value for the whole run, or if the $r$-band
    seeing is worse than 3 arcseconds, or if the operational database
    quality is {\tt BAD}, {\tt MISSING}, or {\tt 
    HOLE}, the field in question is downgraded 
        to BAD (listed as {\bf 0}). 
  \item If the absolute median PSF-aperture difference is smaller than 0.03
        magnitudes in all five bands, and if the absolute value
        of the $s$ color median is smaller than 0.03 mag, and if the
        $s$ color distribution width is smaller than twice
        its median value for the whole run, the field in question is
        upgraded to GOOD (listed as {\bf 2}).
  \item If the median PSF-aperture difference is smaller than 0.02
        magnitudes in all five bands, and if the absolute value
        of the $s$ color median is smaller than 0.02 mag, and if the
        $s$ color distribution width is smaller than 1.5 times
        its median value for the whole run, and the $r$-band seeing
        is better than 2 arcseconds, the field in question is
        upgraded to EXCELLENT (listed as {\bf 3}).
\end{itemize}

In the DR3, 58\% of fields are EXCELLENT, 26\% are GOOD, 13\% are
ACCEPTABLE, and only 3\% are BAD. 

Examples of how to use the RunQA table can be found by selecting this 
table in the CAS Schema Browser on the SkyServer website.

\section{Looking to the Future}
  It is worth emphasizing that doing statistical analyses off the SDSS
  spectroscopic data requires detailed understanding of the
  completeness of any particular sample.  With this in mind, we have
  undertaken various efforts to compile complete samples of various
  subsets of spectroscopic objects, including quasars (e.g., Schneider
  \etal\ 2003), white dwarfs (Kleinman \etal\ 2004), asteroids
  (Ivezi\'c \etal\ 2002), clusters of galaxies (Miller \etal\ 2004)
  and galaxies themselves
  (Blanton \etal\ 2004; this latter includes detailed information on
  the geometry of the sample), as well as photometric samples of
  quasars (Richards \etal\ 2004).  We are actively preparing updated
  samples for DR3.  

\begin{figure}[t]
\centering\includegraphics[width=12cm]{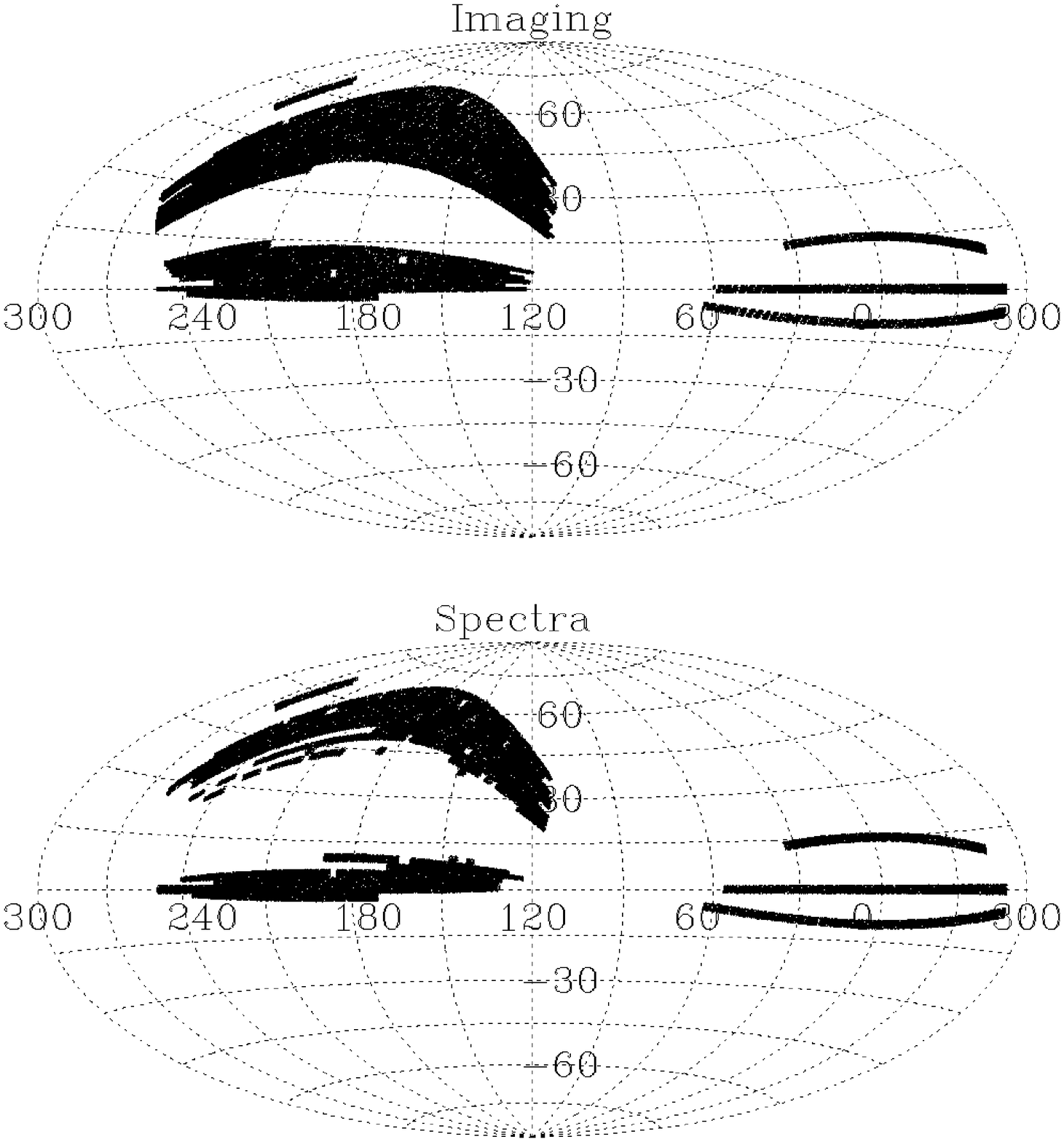}
\caption{The footprint of the SDSS imaging (top) and spectroscopic
  (bottom) data included in DR3.  The former covers 5282 deg$^2$,
  while the latter is 4188 deg$^2$.  The figure is an Aitoff
  projection in equatorial coordinates.  Note that it wraps at $\alpha
  = 300^\circ = 20^{\rm h}$.  The data in the Southern Galactic Cap ($60^\circ
  > \alpha > 300^\circ$) consist of three stripes.  In the Northern
  Galactic Cap, the SDSS is working North from the Celestial Equator,
  and South from a region centered on $\delta \approx +45^\circ$. 
\label{fig:skydist}}
\end{figure}

  Our next data release after DR3 will consist of data taken through
July 2004; it will occur in July 2005.  SDSS survey operations
will end at about that time, and a final data release (DR5) is planned
for early 2006.  As Figure~\ref{fig:skydist} implies, there will still
be a substantial gap between the Northern and Southern pieces of the
sky covered in the North Galactic Cap (i.e., $110^\circ < \alpha <
270^\circ$), and we are actively seeking funds to extend operations
beyond summer 2005 to fill the gap. 

Funding for the creation and distribution of the SDSS Archive has been
provided by the Alfred P. Sloan Foundation, the Participating
Institutions, the National Aeronautics and Space Administration, the
National Science Foundation, the U.S. Department of Energy, the
Japanese Monbukagakusho, and the Max Planck Society. The SDSS Web site
is http://www.sdss.org/. 

\acknowledgements
The SDSS is managed by the Astrophysical Research Consortium (ARC) for
the Participating Institutions. The Participating Institutions are The
University of Chicago, Fermilab, the Institute for Advanced Study, the
Japan Participation Group, The Johns Hopkins University, the Korean
Scientist Group, Los Alamos National Laboratory, the
Max-Planck-Institute for Astronomy (MPIA), the Max-Planck-Institute
for Astrophysics (MPA), New Mexico State University, University of
Pittsburgh, Princeton University, the United States Naval Observatory,
and the University of Washington.

\begin{deluxetable}{lr}
\tablecaption{Characteristics of the SDSS Third Data
  Release (DR3)\label{tab:characteristics}}

\startdata

\cutinhead{\bf Imaging} 

 Footprint area & 5282\ deg$^2$ \\
 Imaging catalog & 141 million unique objects \\
 Magnitude limits:\tablenotemark{a} \\
\qquad $u$      & 22.0 \\
\qquad $g$      & 22.2 \\
\qquad $r$      & 22.2 \\
\qquad $i$      & 21.3 \\
\qquad $z$      & 20.5 \\
Median PSF width     & $1.4^{\prime\prime}$ in $r$ \\
 RMS photometric calibration errors: \\
\qquad   $r$ & 2\% \\
\qquad   $u-g$ & 3\% \\
\qquad   $g-r$ & 2\% \\
\qquad   $r-i$ & 2\% \\
\qquad   $i-z$ & 3\% \\
 Astrometry    & $< 0.1^{\prime\prime}$ rms absolute per coordinate \\

\cutinhead{\bf Spectroscopy}

Footprint area  & 4188\ deg$^2$\\
Wavelength coverage & 3800--9200\AA\\
 Resolution $\lambda/\Delta \lambda$    &1800--2100 \\
 Signal-to-noise ratio &$>4$ per $\sim 1$\AA\ pixel at $g=20.2$\\
 Wavelength calibration & $<5$ km sec$^{-1}$\\
 Redshift accuracy & 30 km sec$^{-1}$ rms for Main galaxies\\
                   & $\sim 99\%$ of classifications and redshifts are reliable\\
 Number of spectra& 528,640\\
 \qquad Galaxies   &                   374,767 \\
 \qquad  Quasars    & 51,027 \\
 \qquad  Stars      & 71,174 \\
 \qquad  Sky        & 26,819 \\
 \qquad  Unclassifiable & 4,853 \\ 
\enddata

\tablenotetext{a}{95\% completeness for point sources in
  typical seeing; 50\% completeness numbers are typically 0.4 mag
fainter (DR1 paper).}
\end{deluxetable}

\end{document}